# Liquid-Metal-Enabled Synthesis of Aluminum-Containing III-Nitrides by Plasma-Assisted Molecular Beam Epitaxy

Yu-Han Liang, T. Nuhfer, and Elias Towe

*Carnegie Mellon University, Pittsburgh, Pennsylvania 15213, USA*

Electronic mail: yuhanlia@andrew.cmu.edu, towe@cmu.edu

## Abstract

Nitride films are promising for advanced optoelectronic and electronic device applications. However, some challenges continue to impede development of high aluminum-containing devices. The two major difficulties are growth of high crystalline quality films with aluminum-rich compositions, and efficiently doping such films p-type. These problems have severely limited use of aluminum-rich nitride films grown by molecular beam epitaxy. A way around these problems is through use of a liquid-metal-enabled approach to molecular beam epitaxy. Although the presence of a liquid metal layer at the growth front is reminiscent of conventional liquid phase epitaxy, this approach is different in its details. Conventional liquid epitaxy is a near-thermodynamic equilibrium process which liquid-metal assisted molecular beam epitaxy is not. Growth of aluminum-rich nitrides is primarily driven by the kinetics of the molecular vapor fluxes, and the surface diffusion of adatoms through a liquid metal layer before incorporation.

   This paper reports on growth of high crystalline quality and highly doped aluminum-containing nitride films. Measured optical and electrical characterization data show that the approach is viable for growth of atomically smooth aluminum-containing nitride heterostructures. Extremely high p-type doping of up to $6 \times 10^{17}$ cm$^{-3}$ and n-type doping of up to $1 \times 10^{20}$ cm$^{-3}$ in Al$_{0.7}$Ga$_{0.3}$N films was achieved. Use of these metal-rich conditions is expected to have a significant impact on high efficiency and high power optoelectronic and electronic devices that require both high crystalline quality and highly doped (Al,Ga)N films.

## I. INTRODUCTION

Aluminum-containing wide-band-gap III-nitride materials provide several practical benefits for ultraviolet photonic devices such as laser diodes,[1,2] light-emitting diodes,[3–5] and advanced photovoltaics.[6,7] However, most aluminum-containing nitride films grown by molecular beam epitaxy still suffer from poor crystalline quality and low doping efficiency; these problems hamper development of high performance devices. Optimal growth conditions for high crystalline quality aluminum-containing nitride compounds generally require growth under group-III-rich conditions. On the other hand, to achieve high p-type doping efficiency, without compensating donor defects, requires nitrogen-rich conditions.[8] These conditions are mutually exclusive.

Many studies have now established that aluminum-containing nitride films with smooth surfaces are best grown under metal-rich conditions; nitrogen-rich conditions lead to rough and faceted surface morphologies that are unsuitable for optoelectronic devices. Neugebauer *et al.,* for example, used analytical arguments from density-functional theory and scanning-tunneling microscopy experiments to demonstrate that smooth nitride films can be achieved under metal-rich growth conditions. Their arguments support the idea that a thin liquid metal layer at the growth front acts as a surfactant that opens up an efficient diffusion channel for lateral adatom transport; this promotes growth of atomically smooth nitride films.[9] Similarly, numerous experimental studies support the notion that presence of excess Ga, which can act as a surfactant, leads to layer-by-layer growth of smooth GaN films.[10,11] A negative aspect of the excess liquid Ga is formation of metal droplets; this prevents creation of abrupt heteroepitaxial interfaces in multilayer structures.

Another challenge that has limited development of high performance ultraviolet light sources is low doping efficiency in high-aluminum-containing nitride films. P-type doping of AlGaN films is complicated by the high activation energy of Mg acceptors in the films. This ranges from 170 meV for x=0, to 510 meV for x=1;[12,13] the low incorporation of Mg dopant,[14] and the significant tendency for self-compensation is attributed to a presence of donor-like native defects or complexes.[15] It is not yet a settled matter what the optimal MBE growth conditions are for synthesis of aluminum-containing III-nitride films with high p-type doping. Most researchers agree that nitrogen-rich conditions are necessary for p-type doping because of the availability of Ga substitutional sites for Mg and the decrease of compensating nitrogen vacancies. The problem however is that under nitrogen-rich conditions, Mg-doped films are likely to be of poor crystalline quality, with rough surfaces. These twin problems make it difficult to synthesize aluminum-rich nitride films that are doped with a high concentration of Mg acceptor impurities and are also of high crystalline quality; overall, there are still no detailed studies on highly doped p-type aluminum-rich AlGaN grown by molecular beam epitaxy.

Recently, Moustakas et al. reported experiments and some theoretical analyses of high quality AlGaN/AlN multi-quantum wells with high internal quantum efficiency. They also reported results on GaN films with high p-type doping concentrations of up to $3\times10^{18}$ cm$^{-3}$. All their samples were grown by plasma-assisted molecular beam epitaxy at extremely high III/V ratios.[16] They suggested that the growth mechanism for their (Al, Ga)N films under the extreme metal-rich conditions was similar to that in conventional liquid phase epitaxy; in effect, the growth could be treated as if it were near thermodynamic equilibrium. Hoke et al. provided additional purely thermodynamic analyses to explain the chemical reactions and the incorporation behavior of group III species and impurities in nitride films under metal-rich growth conditions.[17] They argued that $Mg_3N_2$, one of the by-products of the reaction, is a stable phase in metal-rich conditions. However, this appears to contradict available experimental evidence for molecular beam epitaxy. In fact, there are reports in the literature indicating that $Mg_3N_2$, which causes polarity inversion in III-nitrides, is likely to be formed when there is a high flux of active nitrogen instead.[18] It appears, therefore, that the mere presence of liquid metal during molecular beam epitaxy cannot be easily explained within the thermodynamic equilibrium model because of the low growth temperatures involved, which are far below the thermal equilibrium temperatures for III-nitride compounds.

In this paper we suggest an alternative perspective, supported by experimental evidence, to explain the MBE growth of high aluminum-containing films under extreme metal-rich conditions.

## II. Thin film growth mechanisms and associated thermodynamic and kinetic processes

Early quantitative studies of epitaxial thin film growth of semiconductors were largely empirical.[19,20] It was known then that the driving force for thin film growth was primarily dominated by competition between thermodynamics and kinetics. In the following, we compare and contrast the growth mechanisms for liquid phase epitaxy, dominated by thermodynamics, and molecular beam epitaxy, governed primarily by kinetics.

At its simplest, conventional liquid phase epitaxy is crystallization by cooling and solidification from the liquid (molten) state, where the driving force is the temperature gradient of the cooling along a relevant coordinate. Under the right conditions of chemical equilibrium among mixed crystals in a melt, a crystalline solid can form. For a general $A_xB_{1-x}C$ ternary compound formed from group-III elements and dimeric group-V molecules in liquid-solid equilibrium, the essential chemical reaction is

$$A_{(l)} + B_{(l)} + C_{(l)} = A_xB_{1-x}C_{(s)}. \qquad (1)$$

The phase diagrams that describe the equilibrium conditions are generally determined by consideration of the thermodynamics of the binary systems that form the boundaries of the ternary compound.[21] For the canonical ternary under discussion, the binary systems are $AC$, $BC$, and $AB$. Their deviation from ideality is described in terms of the interaction coefficient

$$\kappa = \frac{\Delta G_m^e}{N_{AC}N_{BC}} = \frac{\Delta H_m - T\Delta S_m^e}{N_{AC}N_{BC}}, \qquad (2)$$

where $\Delta G_m^e$ is the excess free energy of mixing—a form of a fundamental thermodynamic relationship; when normalized by the concentrations $N_{AC}$ and $N_{BC}$ of the binaries $AC$ and $BC$ in the solution, it is the interaction coefficient. The equilibrium conditions between the solid and liquid phases can be evaluated by relating the interaction parameter, $\kappa$, to the activity coefficients, $\gamma_i^{s(l)}$, and the chemical potentials, $\mu_i^{s(l)}$ in the solid and liquid phases. For the canonical three-component compound, the activity coefficient for element $A$ is

$$\text{RT} \ln \gamma_A^l = \kappa_{AC}^l (N_C^l)^2 + \kappa_{AB}^l (N_B^l)^2 + (\kappa_{AC}^l + \kappa_{AB}^l - \kappa_{BC}^l) N_B^l N_C^l. \quad (3)$$

Similar equations for elements $B$ and $C$ can be written down as cyclic permutations of the subscripts. The corresponding chemical potential in the liquid is

$$\mu_A^l(T) = \mu_A^{0l}(T) + RT \cdot \ln(\gamma_A^l N_A), \quad (4)$$

and in the solid it is

$$\mu_{AC}^c(T) = \mu_{AC}^{0c}(T) + RT \cdot \ln(x). \quad (5)$$

The superscripts $l$, $0$, and $c$ denote, respectively, the liquid state, pure state, and the crystalline solid state; $x$ denotes the fraction of $AC$ in the compound $(AC)_x(BC)_{1-x}$ which is assumed to form a perfect solid solution. A similar equation can be written down for the $BC$ compound. The chemical potential of the pure compound above can be related to the chemical potentials of its constituents in the liquid state by the Vieland formula:[22,23]

$$\mu_{AC}^{0c}(T) = \mu_A^{sl}(T) + \mu_C^{sl}(T) - \Delta S_{AC}^F(T_{AC}^F - T) - \Delta C_p \left( T_{AC}^F - T - T \cdot \ln\left[\frac{T_{AC}^F}{T}\right] \right), \quad (6)$$

where $\Delta S_{AC}^F$ is the entropy of fusion, $\Delta C_p$ is the difference between the specific heat of the compound and its supercooled liquid, and $sl$ denotes a stoichiometric liquid. At equilibrium, the chemical potentials of the crystalline solid and of the constituent elements in the liquid are related to each other by

$$\mu_{AC}^c(T) = \mu_A^l(T) + \mu_C^l(T), \tag{7}$$

and

$$\mu_{BC}^c(T) = \mu_B^l(T) + \mu_C^l(T). \tag{8}$$

Two basic assumptions undergird the foregoing discussion. The first is that it is possible to dissolve a solute in a liquid solvent; this is the ideality condition. The second is that the processes are near thermodynamic equilibrium. The latter assumption allows one to use Eqs. (4), (5), and (6) in the relations for the chemical potentials in Eqs. (7) and (8), at equilibrium to derive the liquidus surfaces by elimination of $x$ in the resultant equations, and to derive the solidus lines by elimination of the temperature, T. In this type of analysis, one typically neglects the specific heat term.[23]

In the usual arrangement of a liquid phase epitaxy system, constituents diffuse vertically because of a concentration gradient in the liquid solution to the growth front where they are incorporated into the growing crystal. The growth rate is diffusion-limited, which endows the liquid phase epitaxy process with a decided advantage in that extremely high growth rates in the range of 0.1 − 1.0 μm/minute are possible. Such growth rates are much higher than what is possible in the typical molecular beam epitaxy process,[24] where there is no temperature gradient or vertical diffusion of the constituent species to the growth front. Furthermore, the requirement for high solubility of solutes in bulk solvents cannot be satisfied in the MBE process. For growth of GaN, for example, high solubility of nitrogen in liquid gallium would be required. This solubility can be estimated according to[25]

$$\Delta H_F = RT^2 \frac{dln(x)}{dT} \tag{9}$$

where $\Delta H_F$ is the melting enthalpy and $x$ is the mole fraction of nitrogen in liquid gallium. Generally, the solubility of nitrogen in gallium would require elevated temperature and pressure conditions. Because of this, it remains a challenge to grow GaN thin films from liquid solutions.[26,27] Despite the fact that metastable active nitrogen species, such as atomic nitrogen ($N$), ionic nitrogen ($N_2^+$), and metastable molecular nitrogen ($N_2^*$), could improve solubility of nitrogen in liquid Ga,[28] this is irrelevant for the molecular beam epitaxy process in its vacuum and low temperature environment.

In contrast to liquid phase epitaxy, the MBE process is dominated by surface kinetics which involve impinging vapor fluxes, surface diffusion, arrangement and rearrangement of atoms, and desorption. For the nitrides, the growth rate is controlled by the effective arrival rate of the molecular fluxes, substrate temperature, and the interatomic bonding strength.

It may usually appear that liquid-metal-assisted molecular beam epitaxy is similar to conventional liquid phase epitaxy because of the presence of a liquid metal layer on the surface. This is in fact not the case. The metallic film (which is probably only few atomic layers thick) provides a diffusion channel for adatoms which may then find subsurface adsorption sites. In this respect, plasma-assisted molecular beam epitaxy under extreme metal-rich conditions is dominated by adatom kinetics.

## III. EXPERIMENTAL RESULTS AND DISCUSSION

### *A. Film Growth*

Aluminum-containing single-layer nitride films and quantum-well heterostructures were grown by plasma-assisted molecular beam epitaxy. The films were investigated *in-situ* using a 10-kV reflection-high-energy-electron diffraction (RHEED) system. All the AlGaN films were grown under extreme metal-rich conditions at substrate temperatures of 770 °C, well above the Ga decomposition temperature (750 °C).[29] To achieve high crystalline quality epilayers, two special AlN films were grown: the first is a 100-nm AlN buffer layer deposited at a temperature of 800 °C; this is followed by a thin AlN nucleation layer grown at the intermediate temperature of 750 °C before the actual growth of the AlGaN films. Note that the AlN buffer layers were grown under excess aluminum flux, enabling the formation of liquid aluminum at the growth front. During growth of the (Al,Ga)N films, indium flux, which is used as a surfactant, is introduced; this reduces the surface energy and helps to uniformly distribute the liquid metal layer on the growth surface, thus preventing surface roughness of the grown films. Photoluminescence emission spectra of the $Al_xGa_{1-x}N$ films for different Al mole fractions are shown in Figure 1. As expected, the sharp emission peaks shift to shorter wavelengths with increasing Al composition, irrespective of the Ga flux as long as there was excess metal flux. Figure 2 is a schematic illustration of the action of the constituent species during growth, accompanied by the relevant RHEED pattern in (a); the other illustrations, and the related RHEED patterns in (b) and (c), show a model of what is going on subsequent to closure of the constituent shutters. We observe and note that the RHEED patterns of the $Al_xGa_{1-x}N$ films are visibly dim during the growth. It is believed that the "dimness" of the patterns is due to the presence of an extremely thin liquid metal layer on the growing film surface. It is unlikely that a RHEED pattern would be observable if the metal layer was a thick liquid pool at the growth front; the pristine underlying crystalline order would be inaccessible to the surface electron diffraction process; the electron beam is incident at a glancing angle, with a shallow penetration depth of less than a nanometer.[30] When the constituent fluxes are shut off, while the substrate is maintained at the high growth temperature, very bright and streaky RHEED patterns emerge after only a few minutes as a result of re-evaporation of the thin gallium liquid layer from the surface. After this thermal treatment, the grown films were examined under an optical microscope, and were found to be free of any metal droplets on the surface. Figure 3(a) is a schematic illustration of the relative magnitudes of Al, Ga, In, and active nitrogen atomic fluxes during growth of the $Al_xGa_{1-x}N$ films with an Al composition of 42 % and 58 %, respectively. Since the bonding strength of Al-N (11.52 eV/bond) is much higher than that of Ga-N (8.92 eV/bond) and of In-N (7.72 eV/bond),[31] aluminum adatoms were consumed first by the active nitrogen. Any active nitrogen atoms that were not consumed by the Al then react with Ga atoms and the remaining excess Ga and In atoms contribute to the thin liquid layer at the growth surface. As can be seen in Figures 3(b) and (c), the RHEED pattern is brighter for an (Al,Ga)N film with the relatively low Al fraction of 42 % than that for an (Al,Ga)N film with a composition of 58%; this is because of the much thinner liquid layer on the surface of the lower content aluminum. We also note that the growth rate of the liquid-enabled AlGaN films is constant at about 4.2 nm/min, irrespective of the metal fluxes, as long as there was excess metal flux. This growth rate is higher than that encountered in stoichiometric and nitrogen-rich growth conditions, but much lower than what is typical for conventional liquid phase epitaxy. This fact provides evidence that high solubility of solute atoms in extremely thin metal monolayers is not necessary for molecular beam epitaxial growth under metal-rich conditions. Note that this is in contrast to what the situation is in conventional liquid phase epitaxy where high solubility is crucial. It appears therefore that the chemical reaction that takes place when ternary nitrides are grown under liquid metal conditions by plasma-assisted molecular beam epitaxy may be described by

$$Ga_{(l)} + Al_{(g)} + \frac{1}{2}N_{2(g)}^* \rightarrow AlGaN_{(S)}. \tag{10}$$

The thin liquid gallium layer on the surface serves the function of a fluid-like surfactant to improve the lateral diffusion of the impinging vapor fluxes by lowering the diffusion barrier. One can describe the gallium consumption rate during the growth by the first-order rate equation

$$\frac{d\rho}{dt} = -\Gamma_{AlGaN} - \phi_{des} - \frac{dn_{Ga}}{dt}, \tag{11}$$

where $\rho$ is the Ga adatom density, $\Gamma_{AlGaN}$ is the AlGaN growth rate, $\phi_{des}$ is the Ga desorption rate, and $n_{Ga}$ is the remaining liquid Ga on the surface. This rate equation approximation is a consequence of assuming that the growth rate of the AlGaN film is constant irrespective of the gallium flux whenever excess metal conditions prevail; desorption only occurs at the liquid-vapor interface. In short, the extremely thin metal layer, which is incapable of supporting any temperature gradient on the surface, limits the influence of a thermodynamic driving force. The only reasonable assumption therefore is that growth of nitride films under metal-rich conditions by molecular beam epitaxy is dominated by kinetic processes.

We also studied the optical and structural characteristics of $Al_xGa_{1-x}N/Al_yGa_{1-y}N$ multiple quantum-well films grown under such metal-rich growth conditions. Five-period $Al_xGa_{1-x}N/Al_yGa_{1-y}N$ quantum-well structures were grown on AlN buffer layers at the growth temperature of 770 °C. The barrier and well widths were estimated to be about 8 nm and 2.5 nm, respectively. These estimates are based on scanning transmission electron microscopy (STEM) measurements whose micrographs are shown in Figure 5. The schematic structure of the films and their corresponding low-temperature photoluminescence spectra are shown in Figure 4. The strong emission peak at 299 nm in Figure 4(a) originates from the quantum well region of the $Al_{0.7}Ga_{0.3}N/Al_{0.42}Ga_{0.58}N$ structure, while the weaker peak at 262 nm is associated with emission from the barrier region. The emission peak is blue-shifted to 286 nm in Figure 4(b) for the $Al_{0.7}Ga_{0.3}N/Al_{0.58}Ga_{0.42}N$ multi-quantum structure for the higher aluminum composition of the well region as expected. According to Vegard's law, theoretical estimates require higher Al compositions in the AlGaN well regions for the emission peaks to be pushed toward the short wavelength region. Bhattacharyya *et al.* have suggested that this is probably due to compositional inhomogeneity in aluminum-containing films grown under extreme Ga-rich conditions.[32] This is also confirmed by the Z-contrast high angle annular-dark field (HAADF) electron micrograph of the cross-sectional $Al_{0.7}Ga_{0.3}N/Al_{0.58}Ga_{0.42}N$ quantum-well structure. Figure 5(b) of the enlarged image reveals the cluster-like features in the AlGaN well regions. These features could lead to strong localized exciton emission at the minimum of the energy potential, which by analogy is similar to what happens due to compositional inhomogeneity in indium-containing nitrides[33] This compositional inhomogeneity, however, is unlikely to occur from a thermodynamic point of view because there is no miscibility gap in ternary AlGaN compounds; on the other hand, from a growth kinetics point of view, a compositional inhomogeneity could occur due to the difference in adatom mobilities on the growth surface.

## B. Doped AlGaN films

To further investigate the range of possibilities under the liquid-metal-rich conditions, we have grown films with intentional dopant impurities. A number of 900-nm thick GaN films doped with Mg were grown on c-plane sapphire substrate; an initial layer of a certain thickness of GaN was grown without Mg, and then the Mg shutter was opened without growth interruption. The thin, initial, unintentionally doped GaN film was grown to prevent inversion of the surface polarity. All doped films were grown under metal-rich conditions at substrate temperatures of 770 °C. The Ga-cell temperature was kept constant at 895°C during the growth; this corresponds to a Ga beam equivalent pressure (BEP) of $8.3 \times 10^{-8}$ Torr in our system. The Mg-cell temperature was varied from 310 – 380 °C. We emphasize that a thin liquid metal layer was always present on the film surface during the growths in order to enhance the diffusion of impinging adatoms and to avoid polarity inversion. Figure 6 shows the carrier types determined from Hall measurements, and the corresponding RHEED patterns for various Mg-cell temperatures (hence fluxes). All films showed streaky RHEED patterns, indicating that the surface structure was covered with Ga with no $Mg_3N_2$ formed, which would have induced polarity inversion. This further shows that the liquid metal wetting-layer uniformly covers the growth surface. If this were not the case, the RHEED patterns would be spotty because different surface regions would be terminated with Ga atoms while others would be terminated with nitrogen atoms; this would lead to a rough surface morphology and poor crystalline quality. Note that for high Mg fluxes ($T_{Mg} > 370$ °C), the samples become n-type probably due to formation of Mg clusters at interstitial sites that contribute to donor-type defects. This conversion causes wider and streaky RHEED patterns as seen in the insets of Figure 6(a). On the other hand, Figure 6(b) shows that p-type GaN films with high hole concentrations of up to the mid $10^{18}$ cm$^{-3}$ range with low resistivity can be obtained under the liquid metal growth conditions of molecular beam epitaxy. Electrically active high hole concentrations of up to $6 \times 10^{17}$ cm$^{-3}$ were also achieved in AlGaN films with Al compositions of up to 70 %. The corresponding secondary ion mass spectrometry (SIMS) result shown in Figure 7 indicates that high Mg doping concentrations of up to $5 \times 10^{19}$ cm$^{-3}$ can be achieved in the p-type $Al_{0.7}Ga_{0.3}N$ film. There are a number of reasonable explanations for the high incorporation efficiency. In one scenario, it could be that due to the presence of the thin liquid metal layer on growth surface, the vapor pressure of the Mg buried by the liquid is negligible at the substrate growth temperatures of molecular beam epitaxy. If the surface liquid layer were not there, Mg would otherwise have a higher probability of re-evaporating, making it difficult to dope aluminum-containing nitrides because of the high vapor pressure of Mg which, at the substrate temperatures of 700 °C, is over 10 mTorr. We would also like to comment that at the growth front, the interface between solid and liquid is at a *kinetic* equilibrium because of the high substrate temperatures used in these experiments. Incorporated gallium is easily re-dissolved to liquid metal because of the weaker Ga-N bond (compared to the stronger Al-N bond); nitrogen on the other hand is difficult to re-sublimate into vacuum due to the thin liquid metal layer on the surface. We thus speculate that the higher probability of desorption of incorporated gallium generates more group-III vacancies that become available for Mg incorporation. This would lead to fewer nitrogen vacancies ($V_N$) that could act as compensating defects in p-type material when the growth is carried out under extreme metal-rich conditions. With such a high density of Mg atoms incorporated, the Bohr radii of the free holes overlap, allowing the Mg acceptors to interact and form an impurity band rather than a single acceptor level within the band gap. This probably leads to an extended band tail, and thus a lowering of the activation energy as illustrated in Figure 8. It then becomes plausible and practical that high p-type doping of Al-rich nitrides is possible because of the reduction of the activation energy when growth is carried out under liquid metal growth conditions.

Under normal circumstances, it is difficult to achieve highly conductive n-type III-nitride alloys that have a high composition of Al in them. Our p-type results on aluminum-containing alloys, discussed in the foregoing, have motivated us to re-examine the prevailing status of n-type, silicon-doped nitrides. We have grown a series of Si-doped (Al, Ga) N films under various conditions on top of an undoped GaN layer, which in turn, was grown on top of an AlN buffer layer as discussed earlier. These films were grown at several Si-cell temperatures that ranged from 1150 – 1250 °C; streaky RHEED patterns were observed in all cases. We were able to measure high electron concentrations of up to $10^{20}$ cm$^{-3}$ in both GaN and Al-rich AlGaN fims as shown in Figure 9. These data indirectly provide additional evidence that high solubility is not necessary at the extreme metal-rich growth conditions used; this is especially so since silicon exhibits a low solubility of about 2 % at 700 °C in liquid gallium.[34] As already argued in the case for p-type doping, the high n-type concentration of III-nitrides would not be possible if solubility, which is governed by equilibrium thermodynamics, is necessary for liquid-metal-assisted molecular beam epitaxy.

## IV. SUMMARY

In summary, this work has provided experimental evidence to support the notion that growth of III-nitride alloy compounds by plasma-assisted molecular beam epitaxy under extreme metal-rich conditions is dominated by *kinetics*. Under such conditions, excess metal flux(es) form an extremely thin liquid layer at the growth front at high substrate temperatures. The prevailing growth mechanism is unlike that of conventional liquid phase epitaxy, which can usually be satisfactorily explained by appealing to arguments that assume near thermodynamic equilibrium processes.

One of the key observations from this study is that enhancement of surface diffusion through a thin liquid metal layer is essential to achieving smooth and abrupt interfaces in aluminum-containing nitride heterostructures. This approach appears to offer the best resolution to the long-standing difficulty of obtaining highly-doped AlGaN films with high aluminum content. The results of this work show that it is possible to achieve both high hole and electron carrier concentrations in (Al, Ga)N films. It is expected that such high carrier densities should have a significant impact on the development of high power and high efficiency optoelectronic devices that require high aluminum compositions in the nitride films.

## ACKNOWLEDGMENT

This work was partially supported by DARPA, an agency of the US Department of Defense, through the Space and Naval Warfare Systems Center under Grant

(2006).

[34] P. H. Knck and J. Broder, Phys. Rev. Lett. **90**, 521 (1953).

# FIGURES

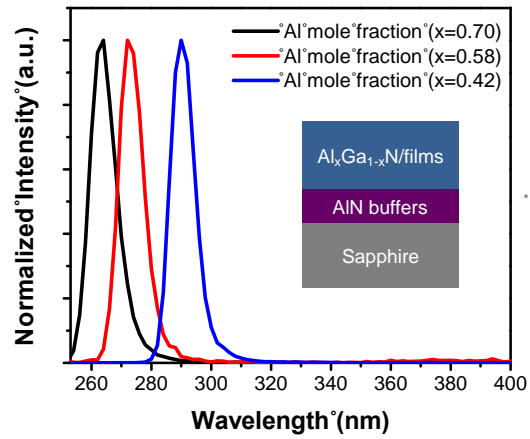

FIG. 1. (Color online) Photoluminescence emission spectra of $Al_xGa_{1-x}N$ films with different Al mole fractions (x).

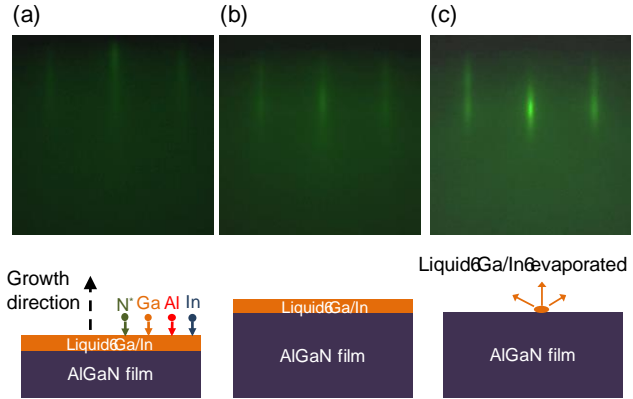

FIG. 2. (Color Online) Illustration of the liquid-metal-enabled epitaxial growth mechanism. Photographs on top show RHEED patterns emanating from an AlGaN film (a) during the growth, (b) 1 minute and (c) 3 minutes after cell shutters were closed, but with the substrate still at the growth temperature.

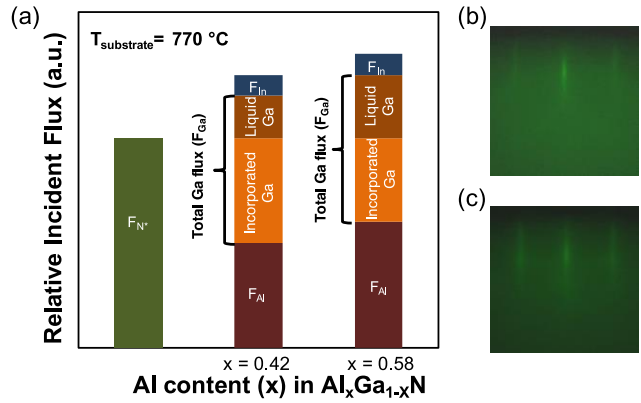

FIG. 3. (Color Online) A schematic illustration of the relative magnitudes of the Al, Ga, In, and active nitrogen fluxes during growth of $Al_xGa_{1-x}N$ films. The RHEED pattern in (b) is for an $Al_xGa_{1-x}N$ film with Al content of 42%, and (c) is for a film with an aluminum content of 58%.

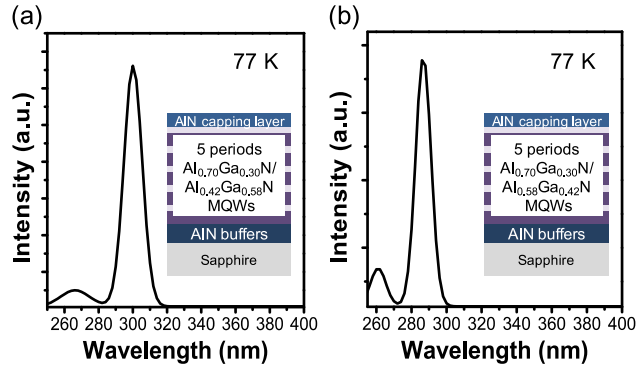

FIG. 4. (Color Online) The schematic structure and photoluminescence emission spectrum for (a) an $Al_{0.70}Ga_{0.30}N/Al_{0.42}Ga_{0.58}N$ multiple quantum-well structure, and (b) for an $Al_{0.70}Ga_{0.30}N/Al_{0.58}Ga_{0.42}N$ multiple quantum-well structure.

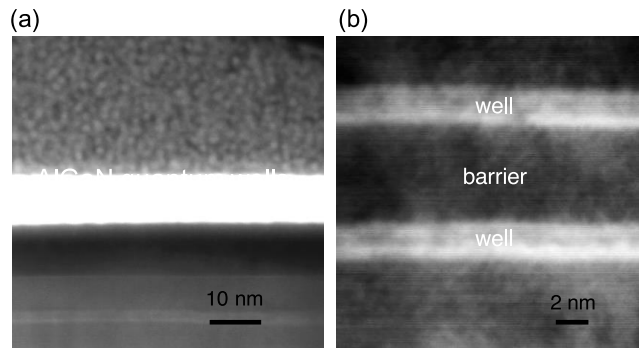

FIG. 5. (Color Online) (a) Cross-sectional Z contrast scanning TEM image of $Al_{0.70}Ga_{0.30}N/Al_{0.58}Ga_{0.42}N$ multiple quantum wells. (b) An enlarged version of the image in (a) reveals the cluster-like compositions in the AlGaN quantum wells.

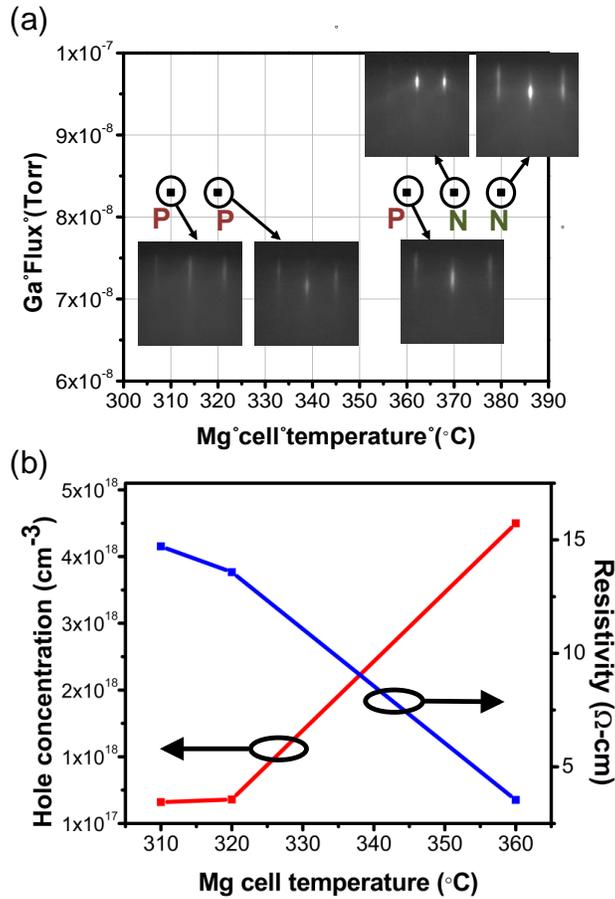

FIG. 6. (Color Online) (a) Carrier type for Mg-doped GaN films for varying Mg-cell temperatures with the corresponding RHEED patterns shown as insets; (b) Resistivity and hole carrier concentration of Mg-doped GaN films for varying Mg-cell temperatures.

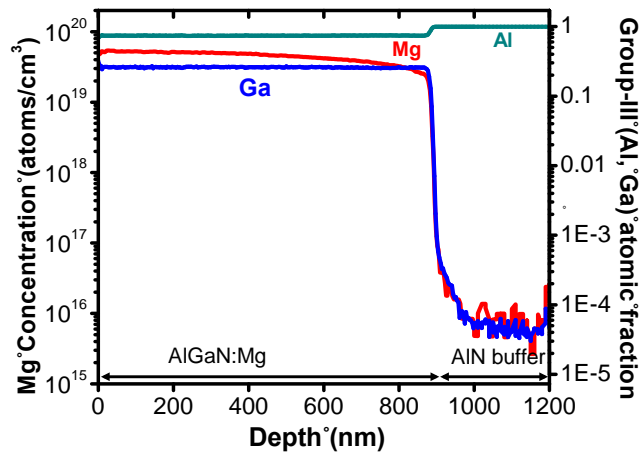

FIG. 7. (Color Online) Secondary ion mass spectroscopy (SIMS) depth profile for a Mg-

doped $Al_{0.7}Ga_{0.3}N$ film.

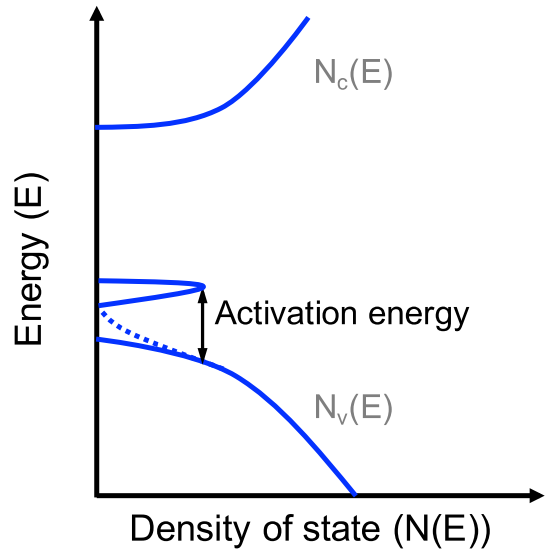

FIG. 8. (Color Online) Band structure of an (Al,Ga)N structure that is highly doped p-type. The dashed line represents extension of a band tail toward the band-edge which tends to lower the activation energy of dopant impurities in aluminum-containing nitrides.

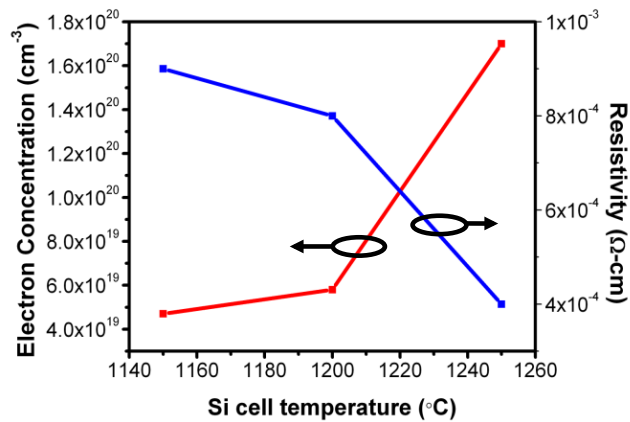

FIG. 9. (Color Online) Electron carrier concentration and resistivity of Si-doped GaN for varying Si-cell temperatures.